\documentclass{article}

\usepackage{arxiv}

\usepackage[utf8]{inputenc}
\usepackage[T1]{fontenc}
\usepackage{hyperref}
\usepackage{url}
\usepackage{booktabs}
\usepackage{amsfonts}
\usepackage{nicefrac}
\usepackage{microtype}
\usepackage{graphicx}
\usepackage{xcolor}
\usepackage{siunitx}
\usepackage{tabularx}
\usepackage{array}
\usepackage{ragged2e}
\usepackage{academicons}
\usepackage{authblk}
\usepackage[numbers]{natbib}
\usepackage{amsmath,amssymb}
\usepackage{multirow}
\usepackage{threeparttable}
\usepackage[justification=justified,singlelinecheck=false]{caption}

\newcolumntype{Y}{>{\RaggedRight\arraybackslash}X}

\title{MARBLE: Multi-Agent Reasoning for Bioinformatics Learning and Evolution}

\author[1,$\dagger$]{Sunghyun Kim}
\author[2,$\dagger$]{Seokwoo Yun}
\author[1,$\dagger$]{Youngseo Yun}
\author[3]{Youngrak Lee}
\author[4,$\ast$]{Sangsoo Lim}

\affil[1]{Division of AI Convergence, Dongguk University, Seoul, South Korea}
\affil[2]{AI Research Team, Ar-ge Inc., Seoul, South Korea}
\affil[3]{Department of Biomedical Engineering, Dongguk University, Goyang-si, Gyeonggi-do, South Korea}
\affil[4]{Department of Computer Science and Artificial Intelligence, Dongguk University, Seoul, South Korea}

\affil[ ]{\vspace{1pt}}
\affil[ ]{\small \textsuperscript{$\dagger$}These authors contributed equally to this work.}
\affil[ ]{\small \textsuperscript{$\ast$}Corresponding author: \texttt{sslim@dgu.ac.kr}}

\date{} 

\begin{document}
\maketitle

\begin{abstract}
\textbf{Motivation:} Developing high-performing bioinformatics models typically requires repeated cycles of hypothesis formulation, architectural redesign, and empirical validation, making progress slow, labor-intensive, and difficult to reproduce. Although recent LLM-based assistants can automate isolated steps, they lack performance-grounded reasoning and stability-aware mechanisms required for reliable, iterative model improvement in bioinformatics workflows.\\
\textbf{Results:} We introduce MARBLE, an execution-stable autonomous model refinement framework for bioinformatics models. MARBLE couples literature-aware reference selection with structured, debate-driven architectural reasoning among role-specialized agents, followed by autonomous execution, evaluation, and memory updates explicitly grounded in empirical performance. Across spatial transcriptomics domain segmentation, drug--target interaction prediction, and drug response prediction, MARBLE consistently achieves sustained performance improvements over strong baselines across multiple refinement cycles, while maintaining high execution robustness and low regression rates. Framework-level analyses demonstrate that structured debate, balanced evidence selection, and performance-grounded memory are critical for stable, repeatable model evolution, rather than single-run or brittle gains.\\
\textbf{Availability:} Source code, data and Supplementary Information are available at \url{https://github.com/PRISM-DGU/MARBLE}.\\
\end{abstract}


\section{Introduction}\label{sec:intro}
Artificial intelligence has become a core methodology in bioinformatics, enabling predictive modeling across diverse applications. Despite substantial progress, the development and refinement of bioinformatics models remain largely manual. Model improvement typically requires repeated cycles of literature review, hypothesis formulation, architectural modification, and experimental validation, demanding significant human expertise and time. Existing bioinformatics AutoML approaches partially alleviate this burden by automating hyperparameter optimization or architecture search; however, they are generally limited to predefined search spaces and lack mechanisms for higher-level scientific reasoning, literature-aware critique, or principled model redesign \citep{barbudo2023eight, he2021automl}.

Recent advances in large language models (LLMs) and agentic AI systems have motivated the development of AI-assisted scientific workflows \citep{wang2023scientific}. Prior studies have explored multi-agent architectures for biomedical discovery \citep{gao2024empowering,schroeder2002multi,huang2025biomni}, autonomous coding agents \citep{trirat2024automl}, and debate-driven scientific reasoning systems \citep{swanson2025virtual,liang2024encouraging}, demonstrating that agent collaboration can support hypothesis generation, planning, and code-level refinement. However, when applied to bioinformatics model development, these approaches remain limited. Many rely on human-in-the-loop supervision \citep{natarajan2025human,mosqueira2023human}, static expert-defined rules \citep{schrittwieser2020mastering}, or task-specific automation, while others focus primarily on executable code correctness driven by programmatic feedback \citep{tang2024biocoder}. Consequently, most existing agent-based systems do not provide an autonomous agentic loop capable of translating insights from the scientific literature into sustained, performance-grounded evolution of model architectures.

Parallel efforts in agent-based neural architecture design and automated research systems further highlight these limitations. Frameworks such as NADER \citep{yang2025nader}, LEMONADE \citep{semnani2025lemonade}, and AlphaEvolve \citep{novikov2025alphaevolve} extend beyond fixed search spaces or manual coding but remain constrained by predefined operator sets, static rulebooks, or machine-gradeable evaluators. 
Similarly, multi-agent scientific reasoning systems—including AI co-scientist \citep{gottweis2025towards} and virtual laboratory frameworks \citep{swanson2025virtual}—leverage literature to ground hypothesis generation and autonomously produce technical artifacts such as experimental protocols or computational scripts. However, these frameworks primarily orchestrate pre-defined models as fixed tools rather than autonomously modifying the internal architectures of the models they employ. 
Collectively, these studies underscore the absence of an end-to-end framework that integrates literature-driven reasoning, structured agent debate, autonomous execution, and performance-based feedback for iterative model refinement in bioinformatics.

To address these limitations, we propose MARBLE (Multi-Agent Reasoning for Bioinformatics Learning and Evolution), a multi-agent meta-reasoning framework for autonomous bioinformatics model refinement. MARBLE targets the iterative evolution of model architectures through structured scientific reasoning grounded in empirical performance, moving beyond prior agentic approaches focused on task-level automation, static architecture search, or code-level correction. The framework follows a closed-loop workflow in which agents perform \emph{Paper Selection}, \emph{Debate-based Ideation}, \emph{Execution and Evaluation}, and \emph{Evolving Memory Update}, with role-specialized researchers, critic, and evaluator agents coordinating under a hybrid modular design to balance reasoning quality and efficiency. We evaluate MARBLE on spatial transcriptomics domain segmentation, drug--target interaction prediction, and drug response prediction, where it consistently improves target model performance and autonomously identifies non-trivial architectural modifications. Although evaluated on specific benchmarks, MARBLE is task-agnostic and broadly applicable to bioinformatics and computational biology modeling problems.

\section{Methods}\label{sec:methods}
\begin{figure*}[!htbp]
  \centering
  \includegraphics[width=\linewidth,keepaspectratio]{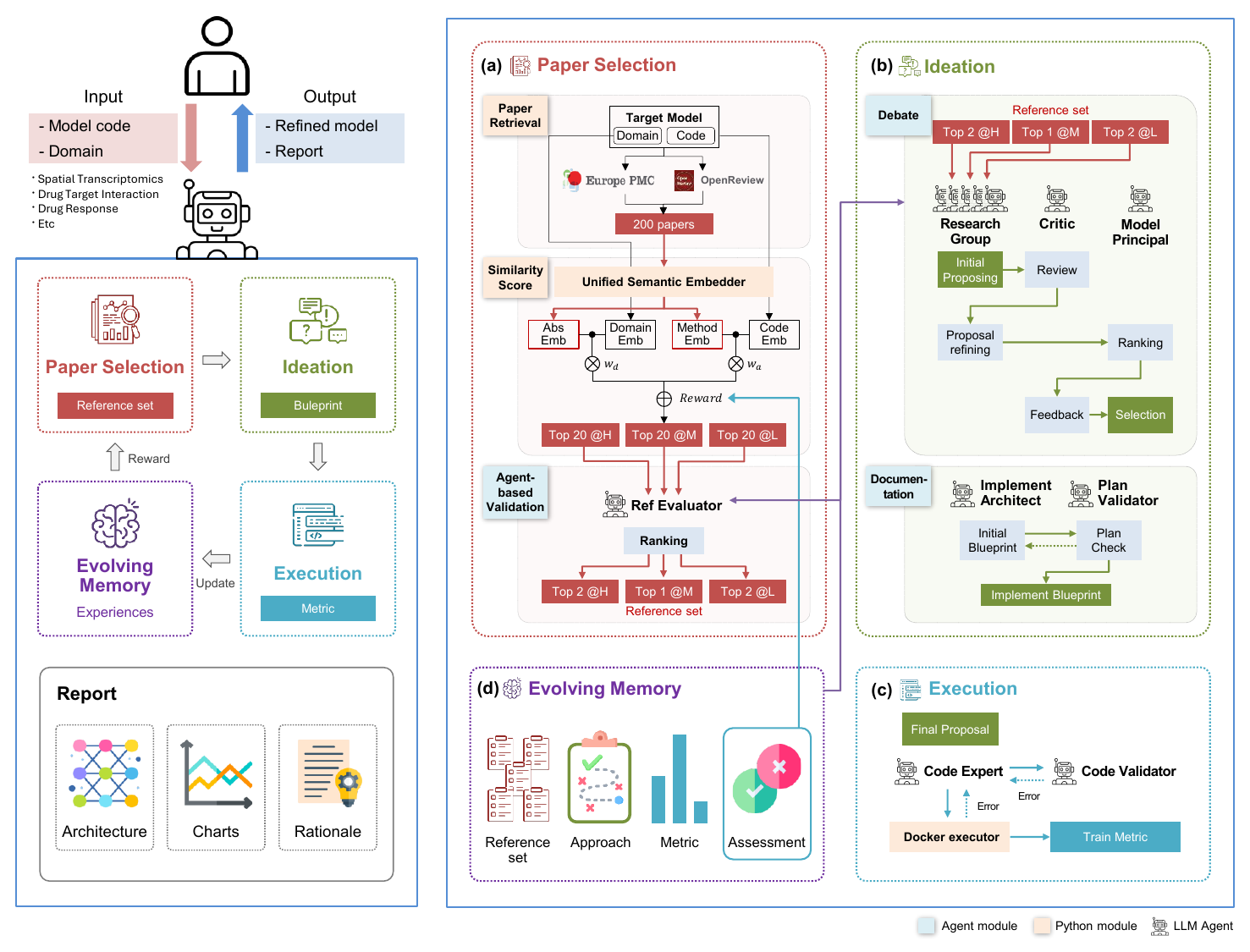}
  \caption{\textbf{Overview of the MARBLE framework.}
  MARBLE is an end-to-end multi-agent system for iterative bioinformatics model refinement. The framework operates in a closed loop comprising \textbf{(a)} literature-aware paper selection, \textbf{(b)} debate-driven ideation among role-specialized agents, \textbf{(c)} autonomous execution and evaluation, and \textbf{(d)} an evolving memory that records outcomes to guide subsequent refinement iterations.}
  \label{fig:overview}
\end{figure*}

We design MARBLE as an end-to-end multi-agent framework for autonomous, iterative refinement of bioinformatics model architectures. Given a target model and task domain, MARBLE performs closed-loop model evolution through coordinated paper selection, debate-driven ideation, execution, and performance-grounded feedback (Figure~\ref{fig:overview}), without task-specific manual intervention after initialization.

\subsection{Paper Selection Module}
The Paper Selection Module (Figure~\ref{fig:overview}a) retrieves and prioritizes scientific literature to guide model refinement. MARBLE adopts a hybrid evaluation strategy that combines embedding-based similarity scoring with agent-based validation, enabling scalable yet robust reference selection.

\subsubsection{Paper Retrieval}
Candidate papers are retrieved using domain- and architecture-specific keywords extracted from the target model. Queries are issued to \emph{Europe PMC} (\url{https://europepmc.org}) and \emph{OpenReview} (\url{https://openreview.net}), and results are filtered to include only Q1 journals or top-tier AI conferences with accessible full text and source code. This process yields a pool of 200 candidate papers per target model.

\subsubsection{Similarity Score}
\begin{figure}[!htbp]
  \centering
  \includegraphics[width=0.7\linewidth,keepaspectratio]{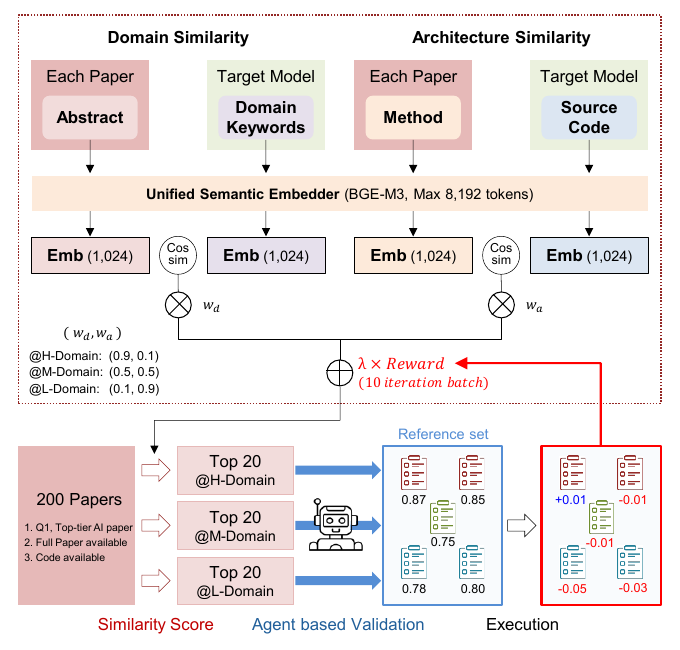}
  \caption{\textbf{Hybrid paper selection with performance-grounded reward update.}
  Candidate papers are scored using embedding-based domain similarity (paper abstract vs.\ target domain keywords) and architecture similarity (paper method vs.\ target model source code) via a unified semantic embedder. The two similarity components are combined using domain and architecture weights $(w_d, w_a)$, which vary across high-, middle-, and low-domain relevance groups, to produce a metric score. From an initial pool of 200 papers, top candidates are filtered per group and further ranked through agent-based validation. Execution outcomes from subsequent refinement iterations generate batch-aggregated reward signals, which update paper selection weights and guide performance-grounded reference prioritization across iterations.
  }
  \label{fig:paper_selection}
\end{figure}

In the first stage of hybrid evaluation, we employ a quantitative scoring mechanism to efficiently process the candidate pool based on semantic relevance and prior utility for an individual paper (Figure~\ref{fig:paper_selection}).

\noindent Unified Semantic Embedder: 
We employ BGE-M3 \citep{chen-etal-2024-m3} as a unified semantic embedder to align heterogeneous scientific text and source code, denoting its dense output as $E$. Its 8,192-token context window and hybrid dense--sparse retrieval enable robust semantic alignment across diverse formats.

\noindent Paper similarity score \& Balancing strategy: 
The total score for an individual paper $p$ at iteration $i$ is defined as the Paper similarity score $S_p^{(i)}$. This score combines the semantic Embedding Similarity $S_{e,p}^{(i)}$ and the batch-aggregated empirical reward $R_p^{(b)}$:
\begin{equation}
    S_p^{(i)} = S_{e,p}^{(i)} + \lambda \cdot R_p^{(b)}
\end{equation}
where $\lambda$ is a weighting factor (set to 0.1) and $b = \lfloor (i-1)/10 \rfloor$ denotes the batch index, grouping every 10 iterations into a single evaluation unit. This empirical reward $R_p^{(b)}$ is introduced to mitigate stochastic noise by reflecting consistent statistical trends rather than transient fluctuations.

The embedding similarity $S_{e,p}^{(i)}$ is calculated as the weighted sum of domain and architecture relevance:
\begin{equation}
    S_{e,p}^{(i)} = w_d \cdot S_{d,p} + w_a \cdot S_{a,p}
\end{equation}
where $S_{d,p}$ and $S_{a,p}$ denote domain and architecture similarity, respectively. To foster diverse ideation \citep{foster2015tradition, musslick2025automating}, we modulate the weights $(w_d, w_a)$ based on the domain category:
\begin{equation}
    (w_d, w_a) = 
    \begin{cases} 
    (0.9, 0.1), & \text{if } \text{H-domain (Exploitation)} \\
    (0.5, 0.5), & \text{if } \text{M-domain} \\
    (0.1, 0.9), & \text{if } \text{L-domain (Exploration)}
    \end{cases}
\end{equation}
For H-domain papers, the system prioritizes contextual stability, while for L-domain papers, it focuses on architectural methodology to inject radical structural priors from distant fields. 
The specific similarity components are calculated as follows. First, the domain similarity $S_{d,p}$ measures the semantic overlap between the target model's domain keyword embedding ($E_d$) and the paper's abstract ($ab$) embedding ($E_{ab,p}$):
\begin{equation}
    S_{d,p} = \text{Sim}(E_d, E_{ab,p})
\end{equation}
Second, to ensure evolutionary stability, the Architecture ($a$) Similarity $S_{a,p}$ incorporates a momentum term. We compare the paper's methodology ($m$) embedding ($E_{m,p}$) against the embeddings of both the historical best and second-best code configurations ($c$):
\begin{equation}
    S_{a,p} = 0.9 \cdot \text{Sim}(E_c^{(i^*_1)}, E_{m,p}) + 0.1 \cdot \text{Sim}(E_c^{(i^*_2)}, E_{m,p})
\end{equation}
where $i^*_1$ and $i^*_2$ denote the iteration indices of the historical best and second-best performing configurations, respectively. 
This dual-anchor comparison prevents the system from overfitting to a single local optimum by anchoring the search to proven architectural priors ($E_c$).

\noindent Batch-aggregated empirical reward: 
The value of $R_p^{(b)}$ is computed by aggregating the counts of success ($N_s$) and failure ($N_f$) relative to the total attempts ($T$) over the cumulative iterations within the current and prior blocks:
\begin{equation}
    R_p^{(b)} = \frac{\sum_{k=1}^{10b} (N_{s,p,k} - N_{f,p,k})}{\sum_{k=1}^{10b} T_{p,k} + 1}
\end{equation}
Based on the final computed $S_p^{(i)}$, we rank the candidate pool and retain the top 20 highest-scoring papers for in-depth qualitative analysis.

\subsubsection{Agent-based Validation}
To complement quantitative metrics, a \emph{Ref Evaluator} qualitatively reviews the pre-screened 20 papers. By analyzing the Methods section for logical feasibility, the agent selects the final reference set: 2 (@H), 1 (@M), and 2 (@L). This ensures a balance between domain stability (Exploitation) and architectural innovation (Exploration).

\subsection{Ideation module}
\label{sec:ideation}

The Ideation module serves as a strategic center of MARBLE (Figure~\ref{fig:overview}b), identifying optimal model modifications through two structured stages: \emph{Debate} and \emph{Documentation}. This process ensures scientific accuracy and computational efficiency while mitigating potential hallucinations \citep{kim2025towards}.

\begin{figure}[!htbp]
  \centering
  \includegraphics[width=0.7\linewidth]{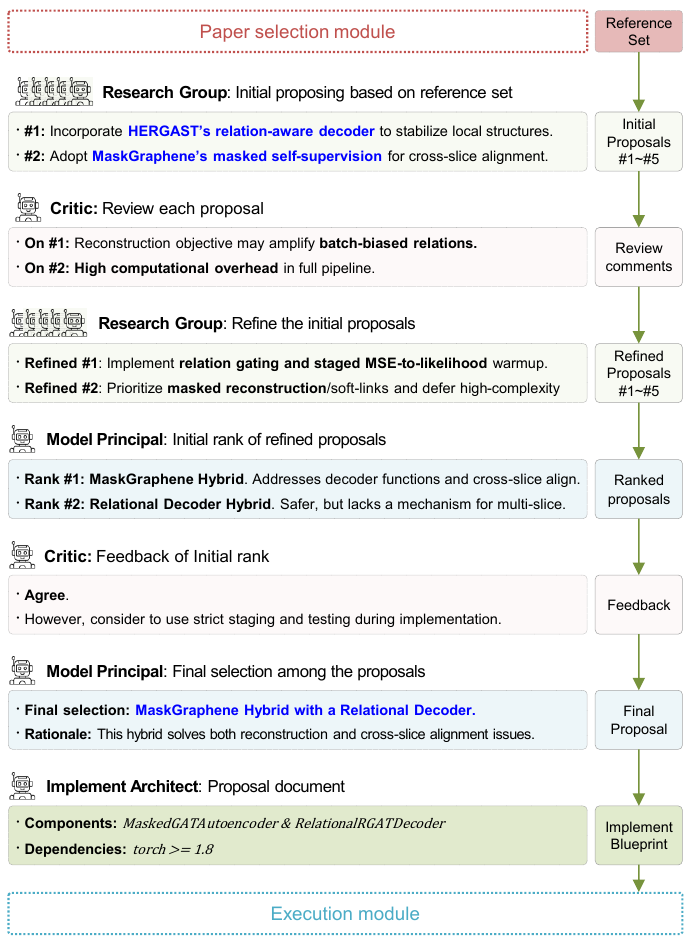}
  \caption{\textbf{Example of debate-driven ideation in MARBLE.}
  Illustrative ideation trace for refining the STAGATE model on a spatial transcriptomics domain segmentation task. \emph{Research Group} generates initial architecture modification proposals based on the selected reference set, which are evaluated and critiqued by a \emph{Critic}. The proposals are then refined and ranked by a \emph{Model Principal}, who incorporates feedback to balance expected performance impact and feasibility, and makes the final selection. Based on this decision, the \emph{Implement Architect} formalizes the chosen approach into a concrete specification for execution.
  }
  \label{fig:ideation}
\end{figure}

\subsubsection{Debate}
The process is initiated by the \emph{Research Group}, which proposes architectural modifications derived from scientific literature. A \emph{Critic} reviews these proposals for compatibility and implementation risks, guiding the group toward refined solutions. Finally, the \emph{Model Principal} ranks these candidates based on their expected performance impact and selects the optimal modification to move forward. Figure~\ref{fig:ideation} illustrates this process.

\subsubsection{Documentation}
The \emph{Implement Architect} formalizes the selected proposal into a technical blueprint by synthesizing relevant code snippets from GitHub repositories. To ensure reliability, a \emph{Plan Validator} performs a final feasibility check. This finalized Implement Blueprint is then dispatched to the Execution Module for autonomous implementation and evaluation.

\subsection{Execution Module}
\label{sec:execution}

The Execution Module (Figure~\ref{fig:overview}c) automates model implementation and evaluation based on the Implement Blueprint. A \emph{Code Expert} integrates the specified architectural changes into the codebase, while a \emph{Code Validator} audits syntactic and configuration integrity. To maximize build success, the module permits up to 10 independent retry attempts for both the validation and isolated Docker execution stages. This process is further robustified by a rebuttal mechanism, allowing the \emph{Code Expert} to contest potential hallucinations of \emph{Code Validator}. Finally, quantitative performance metrics update the Evolving Memory and reference rewards, driving empirical, iterative model evolution. Based on this evolving memory, MARBLE generates an iteration-wise analysis report summarizing architectural and performance changes, presented in Supplementary Figure~\ref{sfig:analysis_report}.

\subsection{Evolving Memory}
\label{sec:memory}

Our Evolving Memory module integrates the Experiential Memory concept \cite{hu2025memory} with the REASONINGBANK framework \citep{ouyang2025reasoningbank}, extending these paradigms beyond internal trajectory distillation to support architectural phenotype refinement through archived structural perturbations derived from external scientific literature.

At each iteration, the module maintains a structured record of the reference set, applied approach, quantitative metrics, and success or failure outcomes, while preserving the historically best-performing iteration as the foundation for subsequent refinement. Every 10 iterations, empirical performance signals are injected as rewards to suppress ineffective references and prioritize promising design directions. By retaining lessons from both successful and failed attempts, the memory steers the system away from redundant errors and toward stable improvement. Through this mechanism, MARBLE enables performance-driven self-evolution informed by empirical evidence, supporting reproducible and stable model refinement across iterations. An illustrative example of iteration-wise records is provided in Supplementary Note~\ref{snote:evolving_memory}.

\subsection{Framework-level Evaluation}
\label{sec:framework_metrics}

We evaluate MARBLE using four framework-level metrics that quantify cumulative performance gains, sustained self-improvement, and execution robustness, independent of task-specific prediction metrics. These metrics adapt success-rate and performance-tracking principles from AutoML-Agent \citep{Trirat2025AutoMLAgent} and iterative progress measures from Coscientist \citep{boiko2023autonomous} to the setting of autonomous bioinformatics model refinement. Details of data preparation for each domain and target model are provided in Supplementary Approach~\ref{sapp:data_preprocessing} and summarized in Supplementary Table~\ref{stab:datasets}.

\subsubsection{Net Performance Gain (NPG)}
NPG measures the cumulative improvement achieved over the refinement process,
\[
\text{NPG} = M_{\text{best}} - M_{\text{initial}},
\]
with the sign interpreted according to whether the task objective is maximization or minimization.

\subsubsection{Normalized Accumulated Improvement over Baseline (NAUI)}
NAUI quantifies the average performance improvement sustained above the baseline across iterations,

\[
\text{NAUI} = \frac{1}{T} \sum_{t=1}^{T} \max\left(0,\; M_t - M_{\text{initial}}\right),
\]

where $T$ denotes the total number of refinement iterations.

\subsubsection{SOTA Improvement Count (SIC)}
SIC counts the number of iterations in which MARBLE attains a new peak performance relative to all previous iterations. For a metric $M_t$ at iteration $t$, an improvement is recorded if
\[
M_t > \max_{i<t} M_i
\]
for maximization objectives (e.g., ARI, AUPRC), or
\[
M_t < \min_{i<t} M_i
\]
for minimization objectives (e.g., RMSE).

\subsubsection{Execution Success Rate (ESR)}
ESR measures execution robustness as the fraction of iterations that successfully complete code integration, training, and evaluation,
\[
\text{ESR} = \frac{N_{\text{success}}}{N_{\text{attempt}}}.
\]

\section{Results}\label{sec:results}

\subsection{Performance across bioinformatics domains} \label{subsec:per}

\begin{figure*}[!htbp]
    \centering
    \includegraphics[width=\textwidth]{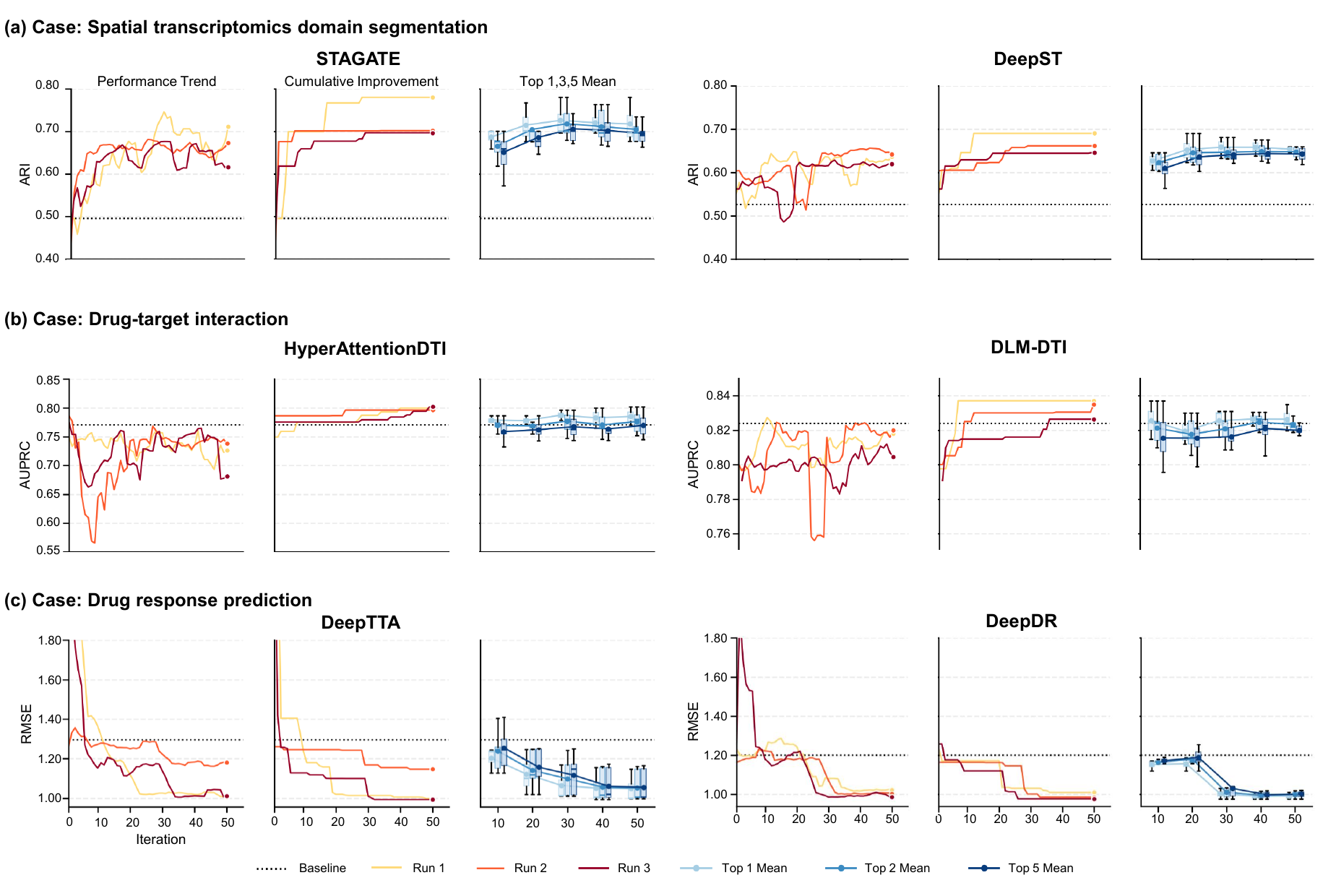}
    \caption{
    \textbf{Iterative performance trajectories of target models refined by MARBLE across bioinformatics domains.}
    Performance trends of MARBLE over refinement iterations for (a) spatial transcriptomics domain segmentation (STAGATE, DeepST; ARI), (b) drug--target interaction prediction (HyperAttentionDTI, DLM-DTI; AUPRC), and (c) drug response prediction (DeepTTA, DeepDR; RMSE). For each model, the left panel shows iteration-wise performance trajectories across independent runs, the middle panel reports cumulative peak performance, and the right panel summarizes mean performance of the top-performing iterations. Across all tasks, MARBLE exhibits repeated performance improvements followed by stable convergence relative to the initial baselines.
    }
    \label{fig:performance}
\end{figure*}

Figure~\ref{fig:performance} summarizes the iterative refinement behavior of MARBLE across six target models spanning three bioinformatics domains. Starting from existing state-of-the-art configurations, MARBLE autonomously refines model architectures through repeated cycles of literature-guided reasoning, execution, and evaluation. Performance is tracked at each iteration and summarized using both iteration-wise trajectories and cumulative best metrics.

Across all models, MARBLE exhibits a consistent refinement pattern characterized by early performance gains followed by stable convergence. Importantly, improvements occur repeatedly across iterations rather than as isolated updates, indicating sustained performance-grounded refinement.

For spatial transcriptomics domain segmentation, MARBLE improves ARI on both STAGATE and DeepST (Figure~\ref{fig:performance}(a)), achieving higher cumulative performance than the original baselines. For drug--target interaction prediction, MARBLE consistently increases AUPRC on DLM-DTI and HyperAttentionDTI (Figure~\ref{fig:performance}(b)). For drug response prediction, MARBLE reduces RMSE on DeepTTA and DeepDR, with most gains occurring in early iterations followed by stable refinement (Figure~\ref{fig:performance}(c)).

Table~\ref{tab:model_performance} summarizes these results using framework-level metrics. MARBLE achieves positive NPG across all six models, non-zero NAUI, multiple SIC events, while maintaining near-perfect ESR. Together, these results indicate that MARBLE provides stable and generalizable iterative refinement across diverse bioinformatics modeling tasks without task-specific manual intervention.

\begin{table*}[htbp]
    \centering
    \caption{\textbf{Framework-level metrics summary across target models.} MARBLE achieves positive NPG across all six models, non-zero NAUI, multiple SIC events, while maintaining near-perfect ESR.}
\label{tab:model_performance}
    
\resizebox{\textwidth}{!}{%
        \begin{tabular}{l |  l c c | c c c c c}
            \toprule
            \textbf{Domain} & \textbf{Target Model} & \textbf{Metric} & \textbf{Baseline} & \textbf{Peak Performance} & \textbf{NPG $\uparrow$} & \textbf{NAUI $\uparrow$} & \textbf{SIC $\uparrow$} & \textbf{ESR $\uparrow$} \\
            \midrule
            \multirow{2}{*}{\textbf{\shortstack[l]{ST}}} 
            & \textbf{STAGATE} & \multirow{2}{*}{ARI $\uparrow$} & 0.496 & 0.726 $\pm$ 0.047 & 0.231 $\pm$ 0.047 & 0.165 $\pm$ 0.007 & 4.333 $\pm$ 1.528 & 0.993 $\pm$ 0.012 \\
            & \textbf{DeepST}  &                      & 0.526 & 0.666 $\pm$ 0.023 & 0.140 $\pm$ 0.023 & 0.047 $\pm$ 0.022 & 5.667 $\pm$ 1.528 & 1.000 $\pm$ 0.000 \\
            \midrule            
            \multirow{2}{*}{\textbf{\shortstack[l]{DTI}}} 
            & \textbf{HyperAttentionDTI} & \multirow{2}{*}{AUPRC $\uparrow$} & 0.771 & 0.799 $\pm$ 0.003 & 0.028 $\pm$ 0.003 & 0.003 $\pm$ 0.004 & 4.667 $\pm$ 2.517 & 1.000 $\pm$ 0.000 \\
            & \textbf{DLM-DTI} &                                  & 0.824 & 0.833 $\pm$ 0.006 & 0.009 $\pm$ 0.006 & 0.014 $\pm$ 0.003 & 5.333 $\pm$ 1.155 & 0.993 $\pm$ 0.012 \\
            \midrule
            \multirow{2}{*}{\textbf{DRP}} 
            & \textbf{DeepTTA} & \multirow{2}{*}{RMSE $\downarrow$} & 1.295 & 1.046 $\pm$ 0.087 & 0.249 $\pm$ 0.087 & 0.938 $\pm$ 0.783 & 10.000 $\pm$ 4.359 & 1.000 $\pm$ 0.000 \\
            & \textbf{DeepDR}  &                       & 1.201 & 0.991 $\pm$ 0.017 & 0.210 $\pm$ 0.017 & 0.123 $\pm$ 0.051 & 5.667 $\pm$ 0.577 & 1.000 $\pm$ 0.000 \\
            \bottomrule
        \end{tabular}%
    }
\end{table*}

\subsection{Framework-level evaluation of MARBLE}
\label{subsec:comparison}

As illustrated in Table~\ref{tab:comparison_function}, while existing agentic frameworks primarily target general-purpose code generation or AutoML, MARBLE is specifically designed for iterative bioinformatics model refinement through a closed-loop, evaluation-driven process. Accordingly, we compare MARBLE with state-of-the-art LLMs and code-generation agents, such as Claude and Codex, which share similar input--output interfaces and represent strong baselines for automated code-level model modification. We evaluate these methods from a framework-level perspective across multiple target models (Table~\ref{tab:comparison}), focusing on whether an agentic system can \emph{reliably and repeatedly} improve model performance while maintaining robust end-to-end execution.

Across all evaluated target models, MARBLE consistently achieves positive and comparatively large NPG, indicating substantial cumulative improvement over the original baselines. In contrast, baseline agents exhibit heterogeneous behavior: some achieve modest gains on a subset of models, while others fail to produce measurable improvements within the same iteration budget. Notably, gains obtained by baseline agents are often confined to a single iteration and do not persist across refinement cycles.

MARBLE further demonstrates superior sustained self-improvement. In addition to consistently positive NPG, MARBLE achieves high NAUI and the highest SIC across target models, indicating that its improvements are repeatedly identified and maintained over successive iterations rather than arising from isolated successes. Throughout this process, MARBLE maintains a near-perfect ESR, reflecting stable execution during continuous refinement. By contrast, baseline agents show fewer improvement events, limited sustained progress, and frequent execution failures, substantially constraining their effectiveness as autonomous refinement frameworks.

Execution robustness further differentiates MARBLE from competing approaches. MARBLE maintains a consistently high ESR across models, indicating reliable code integration, training, and evaluation throughout the refinement process. Several baseline agents, including both code-generation systems and general-purpose LLMs, experience frequent execution failures or incomplete runs, which substantially limits their effectiveness as autonomous refinement frameworks.

Overall, the results in Table~\ref{tab:comparison} show that MARBLE outperforms existing agentic and LLM-based baselines not only in cumulative performance improvement, but also in refinement stability and execution reliability, underscoring the importance of structured multi-agent reasoning, validation, and performance-grounded feedback for robust autonomous model evolution. Details of baseline prompts and execution workflows are provided in Supplementary Note~\ref{snote:baseline_prompts}.

\begin{table}
    \centering
    \caption{\textbf{Comparison of MARBLE with existing coding agents.}}
    \label{tab:comparison_function}
    \resizebox{0.8\columnwidth}{!}{%
    \begin{tabular}{l|c|c|c|c|>{\centering\arraybackslash}p{2.5cm}}
    \toprule
    \textbf{Features} & \textbf{MARBLE} & \textbf{NADER} & \textbf{DeepCode} & \textbf{AutoML-agent} & \textbf{\shortstack{Codex / Claude}} \\
    \midrule
    Main objective & model refinement & model refinement & code generation & AutoML & code generation \\
    \midrule
    Input & code & graph text & code & data & code \\
    Output & code, report & code & code & code & code, report \\
    
    \midrule
    \multicolumn{6}{l}{\textit{\textbf{Paper Selection}}} \\
    Paper retrieval & \textcolor{red}{\scalebox{0.6}{$\bigcirc$}} & \textcolor{red}{\scalebox{0.6}{$\bigcirc$}} & \textcolor{red}{\scalebox{0.6}{$\bigcirc$}} & \textcolor{red}{\scalebox{0.6}{$\bigcirc$}} & \textendash \\
    Similarity score & \textcolor{red}{\scalebox{0.6}{$\bigcirc$}} & \textendash & \textendash & \textendash & \textendash \\
    Agent-based validation & \textcolor{red}{\scalebox{0.6}{$\bigcirc$}} & \textcolor{red}{\scalebox{0.6}{$\bigcirc$}} & \textendash & \textcolor{red}{\scalebox{0.6}{$\bigcirc$}} & \textendash \\
    
    \midrule
    \multicolumn{6}{l}{\textit{\textbf{Ideation}}} \\
    Debate & \textcolor{red}{\scalebox{0.6}{$\bigcirc$}} & \textendash & \textcolor{red}{\scalebox{0.6}{$\bigcirc$}} & \textendash & \textcolor{red}{\scalebox{0.6}{$\bigcirc$}} \\
    Documentation & \textcolor{red}{\scalebox{0.6}{$\bigcirc$}} & \textendash & \textendash & \textendash & \textendash \\
    
    \midrule
    \multicolumn{6}{l}{\textit{\textbf{Execution}}} \\
    Code generation & \textcolor{red}{\scalebox{0.6}{$\bigcirc$}} & \textcolor{red}{\scalebox{0.6}{$\bigcirc$}} & \textcolor{red}{\scalebox{0.6}{$\bigcirc$}} & \textcolor{red}{\scalebox{0.6}{$\bigcirc$}} & \textcolor{red}{\scalebox{0.6}{$\bigcirc$}} \\
    Auto correction & \textcolor{red}{\scalebox{0.6}{$\bigcirc$}} & \textendash & \textcolor{red}{\scalebox{0.6}{$\bigcirc$}} & \textcolor{red}{\scalebox{0.6}{$\bigcirc$}} & \textcolor{red}{\scalebox{0.6}{$\bigcirc$}} \\
    Auto test & \textcolor{red}{\scalebox{0.6}{$\bigcirc$}} & \textendash & \textendash & \textcolor{red}{\scalebox{0.6}{$\bigcirc$}} & \textcolor{red}{\scalebox{0.6}{$\bigcirc$}} \\
    
    \midrule
    \multicolumn{6}{l}{\textit{\textbf{Memory}}} \\
    Storage & \textcolor{red}{\scalebox{0.6}{$\bigcirc$}} & \textcolor{red}{\scalebox{0.6}{$\bigcirc$}} & \textendash & \textcolor{red}{\scalebox{0.6}{$\bigcirc$}} & \textcolor{red}{\scalebox{0.6}{$\bigcirc$}} \\
    Loop (Iteration) & \textcolor{red}{\scalebox{0.6}{$\bigcirc$}} & \textcolor{red}{\scalebox{0.6}{$\bigcirc$}} & \textendash & \textcolor{red}{\scalebox{0.6}{$\bigcirc$}} & \textendash \\
    \bottomrule
    \end{tabular}
    }
\end{table}

\begin{table*}[t]
    \centering
    \caption{
        \textbf{Framework-level metrics comparison of MARBLE against baselines across multiple target models.}
    }
    \label{tab:comparison}
    
    \resizebox{\textwidth}{!}{%
    \setlength{\tabcolsep}{4pt} 
    \renewcommand{\arraystretch}{1.2}
    \begin{tabular}{c c|c|cccc|cccc|cccc|cccc}
    \toprule
    \multicolumn{3}{c|}{\multirow{2}{*}{}} & 
    \multicolumn{8}{c|}{\textbf{Spatial transcriptomics domain segmentation}} & 
    \multicolumn{4}{c|}{\textbf{Drug--target interaction}} & 
    \multicolumn{4}{c}{\textbf{Drug response prediction}} \\
    
    \cmidrule(lr){4-11} \cmidrule(lr){12-15} \cmidrule(lr){16-19}
    
    \multicolumn{3}{c|}{} & 
    \multicolumn{4}{c|}{\textbf{STAGATE}} & 
    \multicolumn{4}{c|}{\textbf{DeepST}} & 
    \multicolumn{4}{c|}{\textbf{HyperAttentionDTI}} & 
    \multicolumn{4}{c}{\textbf{DeepTTA}} \\

    \textbf{Method} & \textbf{Model} & \textbf{Iter.} & 
    \textbf{NPG} & \textbf{NAUI} & \textbf{SIC} & \textbf{ESR} & 
    \textbf{NPG} & \textbf{NAUI} & \textbf{SIC} & \textbf{ESR} & 
    \textbf{NPG} & \textbf{NAUI} & \textbf{SIC} & \textbf{ESR} & 
    \textbf{NPG} & \textbf{NAUI} & \textbf{SIC} & \textbf{ESR} \\
    \midrule
    
    \multirow{2}{*}{\textbf{Claude}} & Opus 4.5 & 10 
    & 0.007 & 0.002 & \textbf{3.000} & \textbf{1.000} 
    & \underline{0.122} & 0.045 & \underline{2.000} & \underline{0.900} 
    & \underline{0.006} & \underline{0.001} & \textbf{3.000} & \textbf{1.000} 
    & \textbf{0.256} & \textbf{0.148} & \underline{1.000} & \underline{0.900} \\
    
    & Haiku 4.5 & 30 
    & \underline{0.251} & 0.033 & 1.000 & \underline{0.800} 
    & 0.081 & 0.037 & \underline{2.000} & 0.733 
    & \textbf{0.026} & \underline{0.002} & \textbf{7.000} & \underline{0.900} 
    & \underline{0.200} & \underline{0.032} & 3.000 & \underline{0.767} \\ 
    \midrule
    
    \multirow{2}{*}{\textbf{Codex}} & 5.2 High & 10 
    & \textbf{0.234} & \textbf{0.195} & \underline{2.000} & \textbf{1.000} 
    & \textbf{0.133} & \textbf{0.081} & \textbf{3.000} & \textbf{1.000} 
    & \underline{0.006} & \textbf{0.002} & \underline{1.000} & \textbf{1.000} 
    & \underline{0.187} & 0.023 & 0.000 & 0.800 \\
    
    & 5.1 Mini & 30 
    & 0.218 & \underline{0.063} & \underline{2.000} & 0.567 
    & \underline{0.120} & \underline{0.040} & \underline{2.000} & \underline{0.833} 
    & $-0.002$ & 0.000 & \underline{2.000} & 0.733 
    & 0.080 & 0.005 & \underline{4.000} & 0.700 \\
    \midrule
    
    \multicolumn{2}{c|}{\multirow{3}{*}{\textbf{MARBLE*}}} 
    & 10 
    & \underline{0.206} & \underline{0.147} & \underline{2.000} & \textbf{1.000} 
    & 0.120 & \underline{0.067} & \textbf{3.000} & \textbf{1.000} 
    & \textbf{0.016} & \textbf{0.002} & 0.000 & \textbf{1.000} 
    & 0.167 & \underline{0.080} & \textbf{4.000} & \textbf{1.000} \\
    
    \multicolumn{2}{c|}{} 
    & 30 
    & \textbf{0.285} & \textbf{0.140} & \textbf{4.000} & \textbf{0.967} 
    & \textbf{0.164} & \textbf{0.087} & \textbf{4.000} & \textbf{1.000} 
    & \underline{0.025} & \textbf{0.003} & \underline{2.000} & \textbf{1.000} 
    & \textbf{0.284} & \textbf{1.120} & \textbf{10.000} & \textbf{1.000} \\
    
    \multicolumn{2}{c|}{} 
    & 50 
    & 0.285 & 0.159 & 4.000 & 0.980 
    & 0.164 & 0.092 & 4.000 & 1.000 
    & 0.031 & 0.002 & 5.000 & 1.000 
    & 0.302 & 0.183 & 12.000 & 1.000 \\
    
    \bottomrule
    \multicolumn{19}{l}{\footnotesize * Peak performance was selected from 3 independent runs.}
    \end{tabular}
    }
\end{table*}

\subsection{Ablation studies} \label{subsec:ablation}
Table~\ref{tab:ablation} presents an ablation study of MARBLE on the STAGATE model using four framework-level metrics: NPG, NAUI, SIC, and ESR. We treat component ablations as controlled stress conditions that corrupt feedback signals, reasoning capacity, or execution reliability, allowing us to assess MARBLE's stability and regression behavior under adverse optimization scenarios. The full MARBLE configuration achieves the strongest and most stable refinement, showing positive cumulative gains, repeated improvement events, and high execution success. Constraining the Paper Selection Module, including domain-only, architecture-only, or single-paper selection, reduces cumulative improvement and improvement frequency, highlighting the importance of balanced reference selection.

Ablations of the ideation components, such as debate and agent role specialization, fail to yield executable or performance-improving modifications, and disabling code validation results in unstable refinement. Removing the evolving memory module decreases sustained improvement while largely preserving execution success. Overall, these results indicate that MARBLE's effectiveness arises from the coordinated operation of its core components rather than any single module in isolation. Under evaluator corruption (fixed or removed reward signals) and reasoning degradation, MARBLE maintains near-perfect execution success (ESR $\simeq$ 1.0) but exhibits sharp drops in improvement frequency (SIC) and cumulative gain (NPG), indicating graceful degradation rather than catastrophic failure. Additionally, an evaluation of the impact of reference paper novelty shows that newly introduced unseen references are more likely to drive sustained performance improvements than those previously encountered (Table~\ref{stab:improvement_stats}).

\begin{table}[htbp]
\centering
\caption{Ablation study of STAGATE using framework-level metrics (NPG, NAUI, SIC, ESR). Best and second-best results are shown in bold and underlined.}
\label{tab:ablation}
\renewcommand{\arraystretch}{1.3}
\setlength{\tabcolsep}{4pt}
\resizebox{0.7\columnwidth}{!}{%
\begin{tabular}{lcccc}
\toprule
\textbf{Configuration}
& \textbf{NPG} $\uparrow$
& \textbf{NAUI} $\uparrow$
& \textbf{SIC} $\uparrow$
& \textbf{ESR} $\uparrow$ \\
\midrule
\textbf{MARBLE}
& \textbf{0.231 $\pm$ 0.047} & \textbf{0.165 $\pm$ 0.007} & \textbf{4.333 $\pm$ 1.528} & \underline{0.993 $\pm$ 0.012} \\
\midrule
\textit{Paper Selection} & & & & \\
\quad w/o Sim. Score
& 0.176 $\pm$ 0.028 & 0.100 $\pm$ 0.036 & 2.500 $\pm$ 2.121 & 0.850 $\pm$ 0.071 \\
\quad w/o Agent Val.*
& 0.176 $\pm$ 0.028 & 0.100 $\pm$ 0.036 & 2.500 $\pm$ 2.121 & 0.850 $\pm$ 0.071 \\
\quad Single Paper ($p$)**
& 0.143 $\pm$ 0.038 & 0.123 $\pm$ 0.041 & 0.667 $\pm$ 0.577 & 0.967 $\pm$ 0.058 \\
\quad w/o $p$ @M, @L
& 0.193 $\pm$ 0.055 & 0.095 $\pm$ 0.058 & 3.000 $\pm$ 2.646 & 0.967 $\pm$ 0.058 \\
\quad w/o $p$ @H, @L
& 0.156 $\pm$ 0.085 & 0.064 $\pm$ 0.032 & 2.500 $\pm$ 0.707 & \textbf{1.000 $\pm$ 0.000} \\
\quad w/o $p$ @H, @M
& \underline{0.203 $\pm$ 0.012} & \underline{0.151 $\pm$ 0.023} & 1.000 $\pm$ 0.000 & \textbf{1.000 $\pm$ 0.000} \\
\midrule
\textit{Ideation} & & & & \\
\quad w/o Debate
& 0.195 $\pm$ 0.034 & 0.092 $\pm$ 0.034 & 2.333 $\pm$ 0.577 & \textbf{1.000 $\pm$ 0.000} \\
\quad w/o Critic
& 0.147 $\pm$ 0.064 & 0.079 $\pm$ 0.067 & \underline{3.500 $\pm$ 0.707} & \textbf{1.000 $\pm$ 0.000} \\
\quad w/o Impl.\ Arch.
& 0.191 $\pm$ 0.040 & 0.118 $\pm$ 0.035 & 2.333 $\pm$ 1.155 & \textbf{1.000 $\pm$ 0.000} \\
\quad w/o Plan Val.
& 0.168 $\pm$ 0.027 & 0.079 $\pm$ 0.020 & 3.333 $\pm$ 1.528 & \textbf{1.000 $\pm$ 0.000} \\
\midrule
\textit{Execution} & & & & \\
\quad w/o Code Val.
& 0.151 $\pm$ 0.011 & 0.079 $\pm$ 0.020 & 2.333 $\pm$ 1.155 & 0.967 $\pm$ 0.058 \\
\midrule
\textit{Evolving Memory} & & & & \\
\quad w/o Memory
& 0.144 $\pm$ 0.073 & 0.057 $\pm$ 0.017 & 2.667 $\pm$ 1.155 & 0.980 $\pm$ 0.020 \\
\quad w/o Reward ($r$)
& 0.193 $\pm$ 0.040 & 0.107 $\pm$ 0.031 & 2.333 $\pm$ 0.577 & \textbf{1.000 $\pm$ 0.000} \\
\quad $r = 1$
& 0.106 $\pm$ 0.081 & 0.055 $\pm$ 0.049 & 2.333 $\pm$ 0.577 & \textbf{1.000 $\pm$ 0.000} \\
\bottomrule
\multicolumn{5}{l}{\footnotesize * Selection based on top 20 embedding scores instead of agent ranking.} \\
\multicolumn{5}{l}{\footnotesize ** Selection based on semantic relevance and prior utility from a pool of 200 candidate papers.}
\end{tabular}
}
\end{table}

\subsection{Case studies} \label{sec:case}

We present two case studies to illustrate how MARBLE translates literature-driven reasoning into concrete architectural refinement across different bioinformatics domains (Figures~\ref{fig:case_study_st} and~\ref{fig:case_study_drp}).

\subsubsection{Case: Spatial transcriptomics domain segmentation}
\begin{figure*}[!htbp]
  \centering
  \includegraphics[width=\linewidth,keepaspectratio]{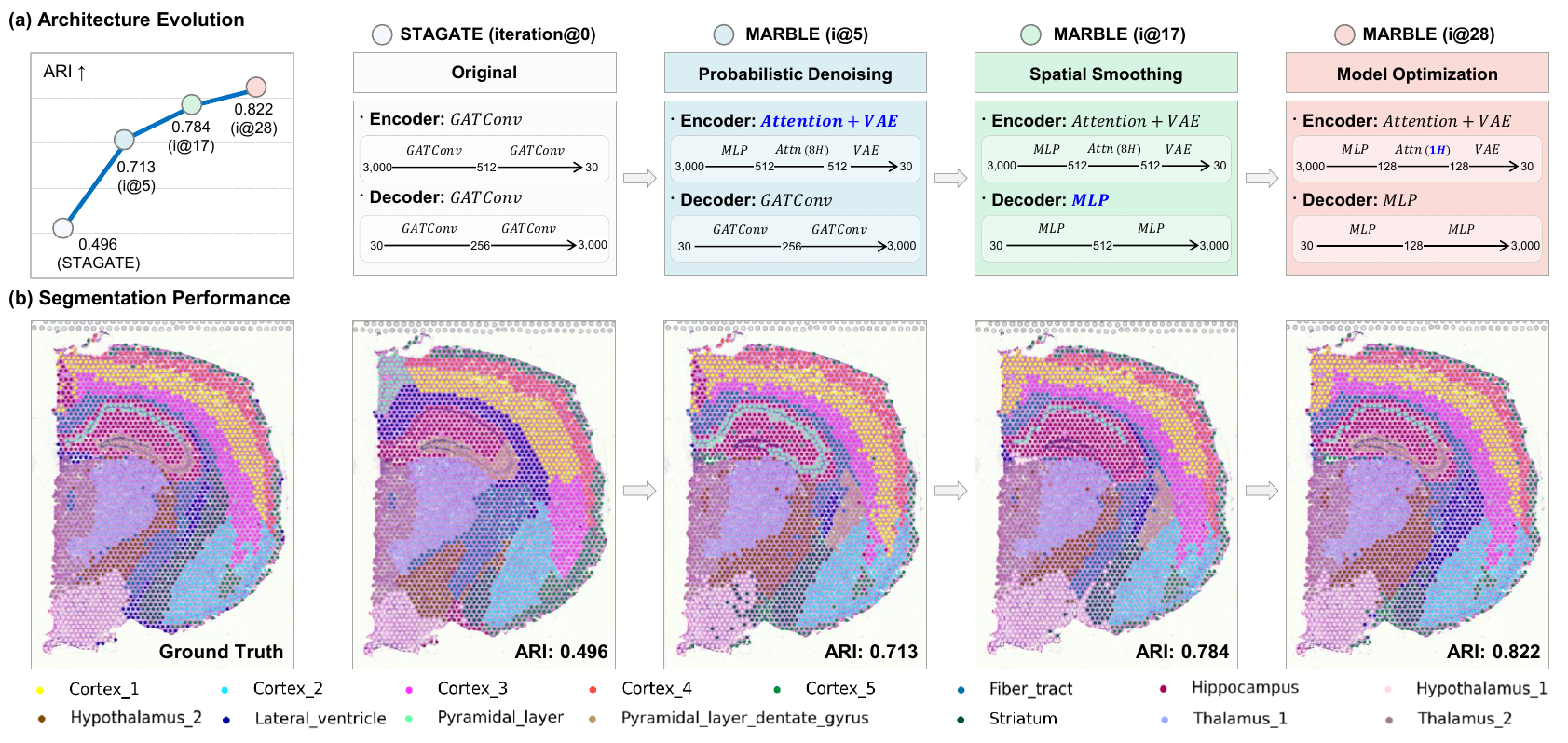}
  \caption{\textbf{Iterative architectural refinement of STAGATE by MARBLE.}
  (a) Performance trajectory (ARI) across refinement iterations and corresponding architectural modifications. Starting from the original STAGATE model, MARBLE successively introduces probabilistic denoising via attention--VAE encoding (i@5), enhances spatial smoothness through decoder redesign (i@17), and performs final architectural optimization (i@28), yielding consistent ARI improvements. (b) Spatial domain segmentation results for each iteration, showing progressively improved spatial continuity and alignment with known anatomical regions.
  }
  \label{fig:case_study_st}
\end{figure*}

Figure~\ref{fig:case_study_st} illustrates the evolutionary refinement of STAGATE for spatial transcriptomics domain segmentation. Starting from the original architecture, MARBLE first introduces VAE-based probabilistic denoising to reduce technical noise and stabilize latent representations. In subsequent iterations, MARBLE expands decoder capacity to enforce spatial smoothness and reduce domain fragmentation. Finally, region-aware multi-task learning is incorporated to capture hierarchical mesoscale structures. These sequential modifications lead to consistent improvements in ARI and produce spatial domain assignments that more closely align with known anatomical organization. The iteration-wise evolution of spatial domain assignments, UMAP-based latent-space projections, and PAGA connectivity graphs during STAGATE refinement is shown in Figure~\ref{sfig:stagate_visualizations}.

\subsubsection{Case: Drug response prediction}
\begin{figure*}[!htbp]
  \centering
  \includegraphics[width=\linewidth,keepaspectratio]{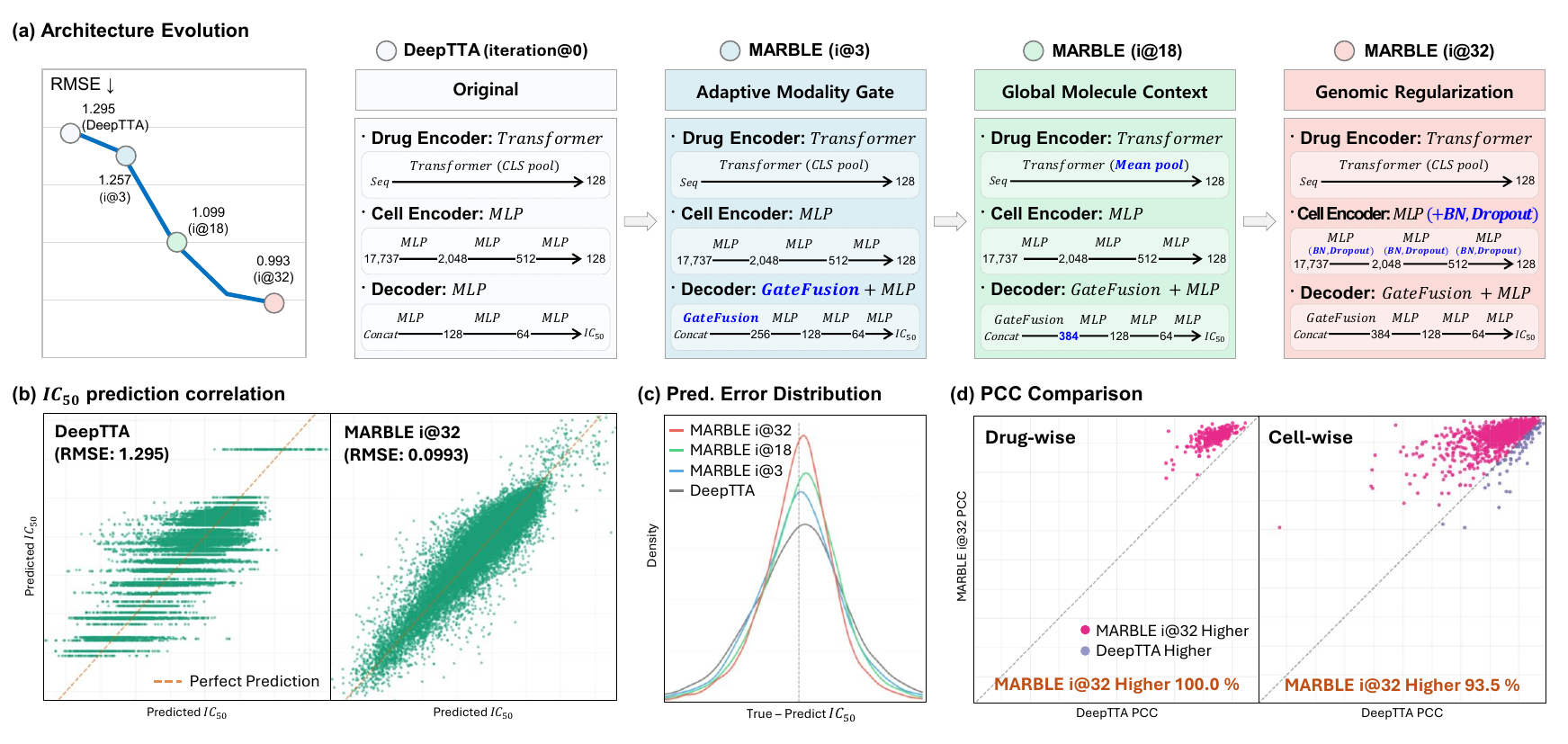}
  \caption{\textbf{Iterative refinement of DeepTTA for drug response prediction by MARBLE.}
  MARBLE progressively refines the DeepTTA architecture across iterations, reducing RMSE through successive modifications, including adaptive modality gating, incorporation of global molecular context, and genomic regularization. The top row shows the RMSE trajectory and corresponding architectural changes at representative iterations. The bottom row visualizes prediction distributions and observed--predicted scatter plots, illustrating improved calibration and reduced prediction error as refinement proceeds.
  }
  \label{fig:case_study_drp}
\end{figure*}

Figure~\ref{fig:case_study_drp} shows the refinement trajectory of DeepTTA for drug response prediction. MARBLE initially introduces adaptive modality gating to better balance drug and cell-line representations. Later iterations integrate global molecular context and genomic regularization to improve generalization. These architectural changes progressively reduce RMSE and yield more stable prediction behavior across iterations. The resulting model demonstrates improved alignment between predicted and observed responses without altering the input modalities.

Together, these case studies demonstrate that MARBLE performs structured, incremental architectural refinement guided by literature evidence and empirical feedback. Rather than relying on isolated modifications, MARBLE accumulates compatible changes across iterations, resulting in stable performance improvements across distinct bioinformatics modeling tasks.

\section{Discussions}\label{sec:discussions}
MARBLE demonstrates that structured multi-agent reasoning grounded in empirical performance can autonomously refine bioinformatics model architectures across diverse tasks. By integrating literature-aware reference selection, debate-driven architectural ideation, reliable execution, and failure-aware memory, the framework achieves consistent and repeatable performance improvements while maintaining high execution robustness. Rather than relying on isolated modifications or manual trial-and-error, MARBLE accumulates compatible architectural changes across iterations, enabling stable model evolution on spatial transcriptomics and drug discovery benchmarks.

MARBLE is designed as research infrastructure rather than a task-specific predictive model. Its primary contribution lies in formalizing model refinement as a closed-loop, performance-grounded process that translates insights from the scientific literature into executable architectural updates, reducing manual effort while preserving transparency and reproducibility.

Several limitations remain. MARBLE currently operates on fixed data modalities, does not introduce new experimental inputs or biological measurements, and is not intended to directly generate novel biological discoveries. In addition, the computational cost of iterative refinement and scalability to substantially larger model classes warrant further investigation, reflecting deliberate design choices that prioritize robustness and generality.

Future work will extend MARBLE to multimodal and multi-omic settings, incorporate resource-aware optimization strategies, and explore tighter integration with experimental feedback loops. More broadly, MARBLE provides a foundation for studying autonomous, performance-grounded model evolution as a systems-level problem in computational biology.

\section{Conclusion}
We presented MARBLE, a multi-agent framework for autonomous, iterative refinement of bioinformatics models through literature-aware reasoning, structured debate, and performance-grounded feedback. Across multiple domains, MARBLE consistently improves model performance while maintaining high execution robustness. By framing model development as a systems-level evolutionary process, MARBLE provides a practical foundation for automating and scaling bioinformatics model design.


\section{Competing interests}
No competing interest is declared.

\section{Author contributions statement}
Conceptualization and original draft preparation were carried out by S.Y. and S.L. Methodology development and software implementation were performed by S.K., S.Y., and Y.Y. Investigation and validation were conducted by S.K., Y.Y., and Y.L. All authors contributed to manuscript review and editing. Supervision and funding acquisition were provided by S.L.

\section{Acknowledgments}
This work was supported by the National Research Foundation of Korea (NRF) grants funded by the Korean government (MSIT) (RS-2025-00560523, RS-2025-18732993, and RS-2022-NR067309). This work was also supported by the Institute of Information \& Communications Technology Planning \& Evaluation (IITP) grants funded by the Korean government (MSIT) under the Artificial Intelligence Convergence Innovation Human Resources Development program (IITP-2025-RS-2023-00254592) and the Information Technology Research Center (ITRC) program (IITP-2025-RS-2020-11201789).

\newcommand{\beginsupplement}{%
  \setcounter{table}{0}%
  \renewcommand{\thetable}{S\arabic{table}}%
  \setcounter{figure}{0}%
  \renewcommand{\thefigure}{S\arabic{figure}}%
  \newcounter{note}
  \renewcommand{\thenote}{S\arabic{note}}
  \newcounter{approach}
  \renewcommand{\theapproach}{S\arabic{approach}}
}

\beginsupplement

\refstepcounter{approach}
\label{sapp:data_preprocessing}

\refstepcounter{table}
\label{stab:datasets}

\refstepcounter{table}
\label{stab:improvement_stats}

\refstepcounter{figure}
\label{sfig:analysis_report}

\refstepcounter{figure}
\label{sfig:stagate_visualizations}

\refstepcounter{note}
\label{snote:evolving_memory}

\refstepcounter{note}
\label{snote:baseline_prompts}

\bibliographystyle{abbrvnat}
\bibliography{references}

\end{document}